\documentstyle[12pt]{article}

\newlength{\extraspace}
\newlength{\extraspaces}
\def\bsklength{.8mm} 

\newcommand{\beq}{\begin{equation}
\addtolength{\abovedisplayskip}{\extraspaces}
\addtolength{\belowdisplayskip}{\extraspaces}
\addtolength{\abovedisplayshortskip}{\extraspace}
\addtolength{\belowdisplayshortskip}{\extraspace}}
\newcommand{\eeq}{\end{equation}}

\newcommand{\beqa}{\begin{eqnarray}
\addtolength{\abovedisplayskip}{\extraspaces}
\addtolength{\belowdisplayskip}{\extraspaces}
\addtolength{\abovedisplayshortskip}{\extraspace}
\addtolength{\belowdisplayshortskip}{\extraspace}}
\newcommand{\eeqa}{\end{eqnarray}}

\newcommand{\newsection}[1]{
\vspace{6mm}
\pagebreak[3]
\addtocounter{section}{1}
\setcounter{equation}{0}
\setcounter{subsection}{0}
\noindent{\large \bf \thesection. #1}
\nopagebreak
\medskip
\nopagebreak}

\newcommand{\newsubsection}[1]{
\vspace{5mm}
\pagebreak[3]

\addtocounter{subsection}{1}
\noindent{ \it \thesubsection. #1}
\nopagebreak
\vspace{2mm}
\nopagebreak}

\newcommand{\ie}{{\it i.e.\ }}

\def\nonu{\nonumber \\[.5mm]}
\def\half{\textstyle{1\over 2}}
\def\pa{\partial}

\def\CD{{\cal D}}

\def\CS{{\cal S}}

\def\CV{{\cal V}}
\def\CW{{\cal W}}
\renewcommand{\hat}{\widehat}
\renewcommand{\tilde}{\widetilde}
\renewcommand{\det}{{\rm det}}

\def\mn{{\mu\nu}}
\newcommand{\e}{{\rm e}}

\newfont{\cmss}{cmss10 scaled\magstep1}
\newfont{\cmsss}{cmss10 }
\def\IZ{\relax\ifmmode\mathchoice
{\hbox{\cmss Z\kern-.4em Z}}{\hbox{\cmss Z\kern-.4em Z}}
{\lower.9pt\hbox{\cmsss Z\kern-.4em Z}}
{\lower1.2pt\hbox{\cmsss Z\kern-.4em Z}}\else{\cmss Z\kern-.4em Z}\fi}
\def\IR{\relax\ifmmode\mathchoice
{\hbox{\cmss I\kern-.5em I}}{\hbox{\cmss R\kern-.5em R}}
{\lower.9pt\hbox{\cmsss I\kern-.5em I}}
{\lower1.2pt\hbox{\cmsss R\kern-.5em R}}\else{\cmss R\kern-.5em R}\fi}

\def\Diff{{\rm Dif\!f}}
\def\SDiff{{\rm SDif\!f}}
\def\diff{{\rm dif\!f}}
\def\sdiff{{\rm sdif\!f}}
\def\lie{\pounds}
\def\delf{\delta_f}
\def\del{\nabla}
\def\d{{\rm d}}

\begin{document}
\setcounter{page}{0}
\addtolength{\baselineskip}{\bsklength}
\thispagestyle{empty}
\renewcommand{\thefootnote}{\fnsymbol{footnote}}	

\begin{flushright}
{\sc MIT-CTP-2462}\\
hep-th/9510147\\
October 1995
\end{flushright}
\vspace{.4cm}

\begin{center}
{\Large
{\bf{Area Preserving Diffeomorphisms and 2-d Gravity}}}\\[1.2cm] 
{\rm HoSeong La}
\footnote{e-mail address: hsla@mitlns.mit.edu\\			
This work is supported in part by funds provided by the U.S. Department of
Energy (D.O.E.) under cooperative research agreement \#DF-FC02-94ER40818.
 }\\[3mm]
{\it Center for Theoretical Physics\\[1mm]			
Laboratory for Nuclear Science\\[1mm]
Massachusetts Institute of Technology\\[1mm]
Cambridge, MA 02139-4307, USA} \\[1.5cm]

{\parbox{14cm}{
\addtolength{\baselineskip}{\bsklength}

Area preserving diffeomorphisms of a 2-d  compact Riemannian manifold with or
without boundary are studied. We find two classes of decompositions of a
Riemannian  metric, namely, h- and g-decomposition, that help to formulate a
gravitational theory which is area preserving  diffeomorphism (SDiff$M$-)
invariant but not necessarily diffeomorphism invariant. The general covariance
of equations of motion of such a theory can be achieved by incorporating
proper Weyl rescaling. The h-decomposition makes the conformal factor of a
metric SDiff$M$-invariant and the rest of the metric invariant under conformal
diffeomorphisms, whilst the g-decomposition makes the conformal factor a
SDiff$M$ scalar and the rest a SDiff$M$ tensor.  Using these, we reformulate
Liouville gravity in SDiff$M$ invariant way. In this context we also further
clarify the dual formulation of Liouville gravity introduced by the author
before, in which the affine spin connection is dual to the Liouville  field.
}
}\\[1.5cm]

{Submitted to {\it Communications in Mathematical Physics}}

\end{center}
\noindent
\vfill


\setcounter{section}{0}
\setcounter{equation}{0}
\setcounter{footnote}{0}
\renewcommand{\thefootnote}{\arabic{footnote}}	
\newcounter{xxx}
\setlength{\parskip}{2mm}
\addtolength{\baselineskip}{\bsklength}
\newpage

\newsection{Introduction}

The geometry of compact oriented manifold is one of the key ingredients  to
study gravitational theories. It is also a very useful tool to investigate
certain two dimensional physics. In traditional approaches of gravity we
require a theory is covariant under diffeomorphisms which are customarily
called the general coordinate transformations with respect to a local
coordinate system. There is additional symmetry in the frame (vielbein) space
which is called the local Lorentz symmetry. In two-dimensions extra information
is needed because we often are led to work with conformal geometry that allows
changes of metric distances, which is a less restrictive geometry compared to
the usual Riemannian geometry. This extra information is provided by Weyl
rescalings which change conformal factor of a metric and are not necessarily
achieved by (conformal) diffeomorphisms. For the Euclidean signature, this is
effectively described by the conformal geometry of Riemann surfaces. Of course,
it is not really essential to require Weyl invariance of a theory,  but 2-d
theories often happen to be Weyl invariant.

The general covariance of a gravitational theory is rooted in the properties of
diffeomorphism group Diff$M$ of manifold $M$ on which the theory is defined and
fundamental variables are covariant objects with respect to Diff$M$.  In this
paper we shall attempt to provide a framework to formulate a theory in which
dynamical variables are not covariant objects with respect to Diff$M$ but
behave covariantly under smaller symmetry: volume preserving  diffeomorphisms.
These are diffeomorphisms that leave a given volume  element invariant and they
form a subgroup SDiff$M$ of Diff$M$.  SDiff$M$ includes isometry group so that
particularly in the flat  case the generators of the Poincar\'e group satisfies
the volume  preserving condition. In 2-d these  are usually called area
preserving  diffeomorphisms for an obvious reason.

The relevance of the area preserving diffeomorphism group SDiff$M$ can be
easily appreciated in the string theory based on the Nambu-Goto
action\cite{ng}, whose lagrangian density is just the area element of a given
surface without specifying any intrinsic metric, so that SDiff$M$ is the
fundamental symmetry for both open and closed strings.   Such lack of manifest
covariance (also the non-linearity of the lagrangian) is usually regarded as a
set-back of Nambu-Goto's approach to string theory.   In the Polyakov string
theory this is  enlarged to the world-sheet Diff$M$ and the Weyl
rescaling\cite{polstr}, nevertheless the area preserving structure resurfaces
in non-critical dimensions through Liouville modes. Furthermore, SDiff$M$
completely excludes any conformal  diffeomorphisms so that one can identify the
genuine dilaton, which in principle is supposed to incorporate all the degrees
of freedom associated with rescaling of metric. Perhaps, this  may indicate
that full understanding of the role of SDiff$M$  might be the key  to
understand the Liouville modes of noncritical string theories as well as the
dilaton. Therefore, it is quite tempting to contemplate on the role of SDiff$M$
more seriously.

In \cite{glg} the author formulates a manifestly area preserving diffeomorphism
invariant gravitational theory, taking analogy of the hydrodynamics of
incompressible, ideal fluids\cite{mw}.  The main idea is to take Diff$M$ as
configuration space of two-dimensional surfaces (as Riemannian manifolds, not
as Riemann  surfaces), then to require SDiff$M$ as underlying symmetry.  This
is different from conventional approaches in which we take the space of  all
metrics with Diff$M$ as underlying symmetry.  This is certainly reasonable at
least on genus zero surfaces because all metrics are related by
diffeomorphisms.  For higher genus surfaces, the space of all metrics is bigger
than Diff$M$ due to the Teichm\"uller deformations, so to apply such a  scheme
it is inevitable to enlarge the configuration space. It turns out that
classically the theory has an equivalent form of action to Liouville gravity as
an induced gravity\cite{polygrav,ddk}, although the underlying structures are
different. Then it is suggested that the quantum theory might be different
because of potentially different quantization due to the area preserving
structure.

We can in fact decompose $\Diff M \simeq \CV\otimes \SDiff M \supset \CW_c
\otimes \SDiff M$ and show that the Liouville gravity action can be rewritten
in terms of $\sigma_\mn = \e^{-2\phi}g_\mn$, where $\phi$ and  $\sigma_\mn$ are
not necessarily covariant objects with respect to Diff$M$,  such that
$\CS_L(\phi,\sigma)$ is SDiff$M$-invariant but not manifestly
Diff$M$-invariant.  Here, $\CV$ is the space of all volume elements with an
equal volume and $\CW_c$ is the space of all conformal diffeomorphisms. It is
also possible that we take SDiff$M$ as underlying symmetry but enlarge the
configuration space to the space of all metrics. Then the approach taken in
\cite{glg} also turns out to provide a dual way of formulating Liouville
gravity with SDiff$M$ as a gauge symmetry. This will be further pursued  in
this paper.

Lately, a certain ``area preserving'' structure in the Liouville gravity was
also investigated in \cite{kmm,jack}.  The main idea is based on the fact that,
from the conformal geometry's point of view, the amount of gauge degrees of
freedom provided by Diff$M$ is equivalent to the amount provided by the
combination of the Weyl rescaling and SDiff$M$. In \cite{kmm,jack} however a
coordinate-choice-dependent (\ie non-covariant) condition of area preserving
structure is used so that it inevitably restricts the jacobian of  any
coordinate change to be unity. In general, the diffeomorphisms whose jacobians
are harmonic transform the Weyl-invariant part of $\sqrt{|g|}R$ still like a
scalar density, so  it is still true if the jacobian is a harmonic function.
However, as a result, the action is not a well-defined integration on a
manifold because it depends on a local coordinate basis.   In general, a
noncovariant condition cannot be imposed globally on a curved manifold. In this
paper we shall find that in fact a covariant condition can be imposed and a
similar argument can be still followed. The key observation is that the
conformal diffeomorphisms $\CW_c$ can be trade off with Weyl rescalings. The
integration can be well defined in terms of a gauge transformation whenever it
is necessary and equations of motion.  This is because a noncovariant object is
usually not globally defined and it depends on a gauge.

In general, we can think of two different cases of SDiff$M$-invariant theories:
First, a lagrangian is written in terms of usual covariant objects and it is
defined on a  manifold with boundary.  Second, a lagrangian  is not written in
terms of covariant objects and is defined on a manifold with or without
boundary. The key idea is that fundamental fields are no longer covariant
objects under Diff$M$, but they are covariant with respect to SDiff$M$,
although not necessarily globally defined over $M$. In both cases action must
be a well-defined integration on $M$ at most up to a gauge transformation with
respect to changes of local coordinate basis.

The first case is a very modest modification. If $\pa M = 0$, $\int_M\hat R$ is
invariant under Diff$M$, being a topological invariant. However, if $\pa M\neq
0$, $\int_M\hat R$ is no longer invariant under Diff$M$ but picks up a boundary
term, although it still is independent from the choice of local coordinate
basis. When we have to deal with  a gravitational theory in which such boundary
is relevant, we are required to introduce an extra surface term that normally
depends on the extrinsic  (geodesic) curvature to preserve the diffeomorphism
invariance  of the theory. A variation of metric induces a variation of  a
surface term which is required to vanish to derive equations of motion. So,
strictly speaking,  boundary terms modify equations of motion too. Thus an
explicit boundary term  is added to derive the same equations of motion as in
the case without boundary, if a gravitational theory is defined on a  manifold
with boundary.   In fact quantum gravity in the path integral formalism is
usually such a case\cite{hawk}.  Suppose we required that a theory were only
invariant under SDiff$M$, we  would not be obliged to introduce such a surface
term because the boundary term actually vanishes. Thus we could speculate that,
if physics near a boundary might break the general covariance, we reduce the
symmetry  to a smaller SDiff$M$ covariance, instead of introducing  a surface
term to restore the general covariance everywhere.

In fact this is quite generic. Any theory defined on a curved manifold can be
interpreted this way. Either we introduce a surface term to preserve the
general covariance, or we could restrict to the volume preserving
diffeomorphism invariance. Of course, whether this is a legitimate thing to do
is another question. We have no intention to abandon the principle of general
covariance which we learned from Einstein's theory,  but we can still try to
see how a theory can be formulated without manifest diffeomorphism invariance
everywhere but only with manifest volume preserving diffeomorphism invariance.
It turns out that in certain cases the variational principle based on SDiff$M$
merely changes the cosmological constant due to the Bianchi identity. Thus it
still maintains the general covariance at the  level of equations of motion.
Also in the flat limit it still does not contradict  to the usual Poincar\'e
symmetry because Poincar\'e transformations are isometric and isometric
diffeomorphisms are volume-preserving.

The second case is more drastic. Usually, a lagrangian of this type is written
in terms of a metric-like object that does not transform like a tensor.
Nevertheless, under SDiff$M$ it has a well-defined transformation property and
that the action defined by integrating such a lagrangian over $M$ is
SDiff$M$-invariant. Weyl rescaling in terms of a proper object, we can rewrite
equations of motion in terms of a metric tensor and the general covariance
can be still recovered away from the boundary.

This paper is organized as follows:
In section two some general properties of diffeomorphism group and
volume preserving diffeomorphism group are explained.  The effects of
diffeomorphisms are described in coordinate-independent  (\ie active) way. In
section three we study the role of area preserving diffeomorphisms in the
Liouville gravity. We also clarify the approach in \cite{kmm,jack}.  In section
four a dual formulation of Liouville gravity in terms of frame introduced in
\cite{glg} is further investigated. The gauge fixing condition  of local
Lorentz symmetry in the frame space motivated by the area preserving
diffeomorphism is analyzed in detail. This  gauge fixing condition is the key
to relate to the Liouville gravity.  Also, some remarks on the zeroth
order formalism are given. Finally, in section five we give comments on
the generic structure of gravitational theories with  area preserving
diffeomorphism group and discuss the relevance of the issues presented in this
paper.

\newsection{DiffM and SDiffM}\\[-10mm]

\newsubsection{{\rm DiffM}}

For a compact oriented manifold $M$, the diffeomorphism group
\footnote{We shall be interested in the orientation preserving case only
so that we shall denote $\Diff M$ to be $\Diff^+M$.}, Diff$M$,
is an infinite-dimensional Lie group which consists of
{\bf C}$^\infty$-diffeomorphisms $f:M\to M$\cite{milnor}.

In this paper we shall adopt an {\it active} way of describing diffeomorphisms:
\beq
\label{eqdif}
f^*: g \to g^{(f)}.
\eeq
In terms of local coordinates, particularly the elements connected to the
identity, i.e. $f\in \Diff_0M$, can be written infinitesimally as
\beq
\label{ei}
f^\mu(x^\alpha) = x^\mu + \epsilon v^\mu(x^\alpha),
\eeq
where $v=v^\mu\pa_\mu\in{\rm Vect}M$ is a vector field on $M$ and
$\epsilon$ is an infinitesimal parameter. $v$'s form a Lie algebra so that we
denote $v\in\diff M$. We can express the change of a metric under
$\Diff_0M$ infinitesimally (for example, see \cite{friedan}) as
\beq
\label{eii}
f^*: ds^2=g_\mn dx^\mu dx^\nu \mapsto ds_f^2=g_\mn^{(f)}dx^\mu dx^\nu
=(g_\mn + \delf g_\mn) dx^\mu dx^\nu
=g_\mn(f) df^\mu df^\nu.
\eeq
Then $\delf g_\mn$ is nothing but the Lie derivative defined by
one-parameter subgroup of $\Diff_0M$:
\beq
\label{lieder}
\lie_v g_\mn =\delf g_\mn= \del_\mu v_\nu +\del_\nu v_\mu,
\eeq
where $\del_\mu$ is the covariant derivative with respect to the Riemannian
connection. The tensorial property  of the metric which is also useful in the
{\it passive} approach is given by incorporating metric form changes as well as
coordinate changes on an overlap of coordinate charts such that
\beq\label{eqcoord}
\tilde{g}_\mn(f) {\pa f^\mu\over \pa x^\alpha}{\pa f^\nu\over \pa x^\beta}
=g_{\alpha\beta}.
\eeq
Weyl rescalings are changes of a metric $g \to \e^{2\phi} g$
that are not necessarily accomplished by diffeomorphisms.

More precisely, now let us clarify the relation between the active approach
and the passive approach by explicitly comparing them. Note that in the above
$\tilde{g}_\mn(f)$ is not the same as $g^{(f)}_\mn(x)$. Infinitesimally,
eq.(\ref{eqcoord}) can be expanded according to eq.(\ref{ei}) as
\beq
\label{eqcexp}
\tilde{g}_{\alpha\beta}(x)
+ \epsilon\left(v^\sigma\pa_\sigma\tilde{g}_{\alpha\beta}
+\tilde{g}_{\mu\beta}\pa_\alpha v^\mu
+\tilde{g}_{\mu\alpha}\pa_\beta v^\mu
\right) = g_{\alpha\beta}(x).
\eeq
Since the form change between $\tilde{g}_{\alpha\beta}$ and $g_{\alpha\beta}$
should be of order $\epsilon$ infinitesimally, in the second term of the LHS
we can replace $\tilde{g}_{\alpha\beta}$ with $g_{\alpha\beta}$. Thus we
recover
\beq
\label{eqlie}
g_\mn - \tilde{g}_\mn = \del_\mu v_\nu +\del_\nu v_\mu =
\delf g_\mn = \lie_v g_\mn
\eeq
At least in the leading order of $\epsilon$ the difference between
eq.(\ref{eii}) and eq.(\ref{eqcoord}) is whether one defines in an  {\it
active} way or in a {\it passive} way, which is reflected by the sign of the
change in the metric form.

At this moment we would like to call the reader's attention to the fact that
the coordinate invariance of an object is not necessarily the same as the
diffeomorphism invariance of the object.  If an object satisfies a vanishing
Lie derivative, it is said to be Diff$M$-invariant.  For tensors, this implies
in particular the tensorial form invariance. For example, a metric tensor is
not Diff$M$-invariant, but invariant only under isometries.  In particular, at
the level of equations of motion Diff$M$-invariance is equivalent to the
general covariance. But at the level of an action Diff$M$ invariance is a
stronger statement than  the invariance under coordinate transformations
because  a coordinate transformation is merely a change of coordinate basis. It
is absolutely necessary for an action to be independent from a choice of local
coordinate basis to be a well-defined integration on $M$.

For a scalar $S(x)$ we know that under the general coordinate transformation it
should transform like $\tilde{S}(f) = S(x)$, but  $\lie_v S$ (which can be
expressed locally as $v^\mu \pa_\mu S$) does not necessarily vanish. The former
incorporates coordinate change as well as a form change. However an
infinitesimal diffeomorphism ignores the coordinate change but measures the
form change only. These two are not the same in general for other tensors
either. A lagrangian density has a form of $\sqrt{|g|}S$, where  $g = \det
g_\mn$.   Under coordinate transformation, this transforms like a scalar
density. Under $\Diff_0M$,  $\lie_v (\sqrt{|g|} S) = \pa_\mu (v^\mu \sqrt{|g|}
S)$. These two changes are not the same even infinitesimally because the
jacobian  is given by $J = 1 + \pa_\mu v^\mu$.

Note that the equality between volume elements in two-dimensions
\beq
\label{eqqai}
d^2f \sqrt{|g'(f)|} = d^2x \sqrt{|g(x)|}
\eeq
is always true for any $f$ as a coordinate transformation, which can be  shown
easily using jacobian.  This simply means a volume element does not depend on a
choice of a local coordinate basis and it is also necessarily for a volume
element to be well-defined on a curved manifold. There is another way to to
check this invariance without using the jacobian. The LHS of eq.(\ref{eqqai})
can be expanded in terms of $f = x+v$ as
$$d^2f \sqrt{|g'(f)|} = d^2x \sqrt{|g'(x)|} (1 + \del_\mu v^\mu).$$
We can now recover the identity eq.(\ref{eqqai}) for any $f\in \Diff M$, using
the equality
$$d^2x \sqrt{|g'(x)|} = d^2x \sqrt{|g(x)|}(1-\del_\mu v^\mu).$$
This simple computation without using jacobian is significant in the sense
that all the ingredients are covariantly defined over manifold $M$ without
preferred choice of a coordinate system. Also this is a source to the confusion
that the coordinate invariance is equivalent to the diffeomorphism invariance,
which is not always true. To prove whether an action of noncovariant objects
is a well-defined integration or not, this method is very useful.

We can compare each step to the diffeomorphism case.
Under coordinate transformation $x\to f$ we obtain
$$\sqrt{|g'(f)|} -\sqrt{|g(x)|} = -\sqrt{|g(x)|} \pa_\mu v^\mu .$$
On the other hand, under diffeomorphism $f = x + v$ we have
$$\lie_v\sqrt{|g(x)|} = \sqrt{|g(x)|} \del_\mu v^\mu . $$
Ignoring the sign difference, which merely reflects the active and the passive
way of using the transformation, the difference of the above is
$v^\mu\pa_\mu\sqrt{|g(x)|}$, which precisely measures the functional change
with
respect to the coordinate change. Thus from Diff$M$'s point of view
eq.(\ref{eqqai}) is not a proper way to compare and the jacobian is not a good
object to use either. Upon integration the coordinates are in fact dummy so
that we should really compare $d^2x \sqrt{|g'(x)|}$ and $d^2x \sqrt{|g(x)|}$.
Note that the equality $d^2x \sqrt{|g'(x)|} =d^2x \sqrt{|g(x)|}$  can only be
achieved if $\lie_v\sqrt{|g|} = 0$, \ie\ $\del_\mu v^\mu = 0$,  which is the
condition to define area preserving diffeomorphisms.

\newsubsection{{\rm SDiffM} in general}

Volume preserving diffeomorphisms are not necessarily characterized by the
property of preserving volume itself because diffeomorphisms  also preserve
volume as we pointed out in the previous section.
We need to require a stronger condition to distinguish them.

The {\bf volume preserving conditions} are defined by the follows:
For $f\in\Diff M$ and $v\in \diff M$, \\[-8mm]
\begin{itemize}
\item{}{[{\bf VP1}]\quad $g^\mn\delf g_\mn =0$, i.e. $\del_\mu v^\mu =0$.}
\item{}{[{\bf VP2}]\quad  $v^\mu$ is tangential to the boundary $\pa M$.}
\end{itemize}

If a vector field $v$ satisfies the above conditions, then $v$'s form a Lie
subalgebra $\sdiff M$ of $\diff M$ such that  $[u, v] = w$ and $\del_\mu u^\mu
= 0 = \del_\mu v^\mu$ implies  $\del_\mu w^\mu = 0$. The corresponding $f$ is
called a volume preserving diffeomorphism.  These vector fields generate a
subgroup of $\Diff M$ called volume preserving  diffeomorphism group,
$\SDiff_{\hat\mu} M$ for a volume element $\hat\mu$.  [{\bf VP1}] is the
condition which leaves  this volume form invariant, particularly,
$\delf\sqrt{|g|}=0$  and [{\bf VP2}] prohibits any area change over boundary
from  occurring. In terms of the codifferential $\delta$, [{\bf VP1}] becomes
$\delta\hat v =0$ for the one-form $\hat v=g_\mn v^\mu\d x^\nu$  corresponding
to vector field $v$. The infinitesimal actions of Poincar\'e group actually
satisfy [{\bf VP1}], being isometric.

For any vector $V^\mu$ on $M$, if $f\in \SDiff M$, then
\beq
\label{eapid}
\del_\mu^{(f)}V^\mu =\del_\mu V^\mu,
\eeq
where $\del^{(f)}_\mu$ is a covariant derivative in terms of $g_\mn^{(f)}$.
 This identity is due to
$\delf\Gamma^\alpha_{\alpha\mu} =0$ if $f\in\SDiff M$.

In terms of zweibeins $e_\mu^a$ and their inverses $E_a^\mu$ such that
$$
g_\mn=\eta_{ab}e_\mu^a e_\nu^b,\quad\quad g^\mn=\eta^{ab}E_a^\mu E_b^\nu,
$$
the torsion-free affine spin connection\footnote{This notation can be
potentially confusing in other than two-dimensions. Even in two-dimensions
sometimes we need to keep in mind the hidden frame indices because they
determine the transformation property of the object in the frame space.}
$\omega_\mu\equiv\omega_{\mu\ b}^{\ a}$ is given by
\beq
\label{espc}
\omega_{\mu\ b}^{\ a}=-E^\alpha_b\del_\mu e_\alpha^a.
\eeq
Using $\delf g_\mn=\eta_{ab}(\delf e_\mu^a e_\nu^b + e_\mu^a\delf e_\nu^b)$,
[{\bf VP1}] now reads
\beq
\label{efrap}
g^\mn\delf g_\mn = 2E^\mu_a\delf e_\mu^a =0.
\eeq
This spin connection is not necessarily globally defined over $M$ as
there is no global frame over $M$. As is well known, it quite resembles
a gauge theory.

Just to clarify the notation,
the Lie derivative acting on a p-form $\alpha$ is read as
\beq
\label{elie}
\lie_v\alpha = (\d i_v + i_v\d) \alpha,
\eeq
where $i_v$ is the inner product with respect to a vector field $v$. Since
$i_v$ lowers the rank of a differential form, in particular  $i_v S = 0$ for a
scalar $S$.

In fact in n-dimensions any n-form $\Omega$ satisfies that
\beq
\label{exi}
\int_M\lie_v\Omega = \int_M\d i_v\Omega.
\eeq
Unless $v$ and $\Omega$ are globally defined on M, there is no obvious reason
why $i_v\Omega$ is globally defined. However, let us assume it is globally
defined to apply Stoke's theorem so that
\beq
\label{exii}
\int_M\lie_v\Omega = \int_{\pa M}i_v\Omega.
\eeq
For $v\in\diff M$ this does not necessarily vanish.  However, if $v\in\sdiff
M$, the RHS vanishes because $v\parallel\pa M$. This property of SDiff$M$ is
very important for us to incorporate volume preserving diffeomorphism as a
symmetry of a given physical system  defined on a manifold with boundary.

Finally, we quote one important theorem:
Omori-Ebin-Marsden's theorem\cite{omori,ebmars}. It states that $\Diff M$ is
diffeomorphic to  $\CV\otimes\SDiff_{\hat{\mu}} M$, where $\CV=\{{\hat\nu}\}$
is the space of  all volume elements that satisfy  $\int_M {\hat\nu} =\int_M
{\hat\mu}$. This theorem is not only true for manifold without boundary but
also true for manifold with boundary.

\newsubsection{Metric decompositions}

Notice that conformal Killing vectors do not satisfy the volume preserving
conditions. Using this property, we can separate SDiff$M$ completely from any
conformal transformations.
This can be done by decomposing the metric in following way.
Let $g_\mn = \e^{2\phi} h_\mn$, then we shall call it {\bf h-decomposition}
for future referencing purpose, if
$$\delf g_\mn = \e^{2\phi}\delf  h_\mn +  h_\mn\delf\e^{2\phi} ,$$
where in n-dimensions
\beqa
\label{ehtg}
\delf h_\mn &=& v^\alpha\pa_\alpha h_\mn
		+  h_{\mu\alpha}\pa_\nu v^\alpha
		+  h_{\alpha\nu}\pa_\mu v^\alpha
		- {2\over n}\del^{(h)}_\alpha v^\alpha  h_\mn ,\\[1mm]
\label{ehtge}
\delf\e^{2\phi} &=& {2\over n}\del_\alpha v^\alpha \e^{2\phi}.
\eeqa
For notational convenience we defined $\del^{(h)}_\alpha$ as a covariant
derivative in terms of $h_\mn$, although $h_\mn$ does not necessarily transform
like a metric tensor. Note that $ h^\mn\delf h_\mn = 0$ for any $f\in\Diff M$.
In this sense we can regard $h_\mn$ as sort of a metric to SDiff$M$.  This
decomposition now clearly shows that $ h_\mn$ is invariant under any conformal
diffeomorphisms, whilst $\e^{2\phi}$ is invariant under SDiff$M$. Both
variations can vanish at the  same time only if $f$ is an isometry.  So we have
explicitly separated the conformal factor from the rest and $\phi$ can be
identified as a true dilaton because fixing $\phi$ fixes conformal degrees of
freedom completely\footnote{Fixing these degrees of freedom should generate an
independent scale in a theory. It is shown in \cite{diltp}  that the dilaton
indeed generates a scale which is independent from the Newton's constant  in
terms of its own scale parameter in two-dimensions.}.
Now a Weyl transformation of $g_\mn$ can be regarded as a change
of $\phi$ so that the property of eq.(\ref{ehtge}) can be preserved.

If we have a theory of an action $\CS(h_\mn, \phi, \cdots)$ in which  $h_\mn$
takes the role of a metric tensor and if it is invariant under eqs.(\ref{ehtg})
for $\del_\mu v^\mu = 0$,  then the theory is SDiff$M$-invariant. If $\CS$ does
not contain $\phi$, then it is also Weyl-invariant. A simple example is to take
$\int_M \sqrt{|h|} R(h)$. It does not depend on $\phi$ at all so that we can
say it is not manifestly Diff$M$-invariant.  Nevertheless, it can be shown that
the action does not depend on a choice of local coordinate basis, so it is a
well-defined integration on $M$.  Under SDiff$M$ the action is invariant, if
$v^\mu\pa_\mu\phi = 0$ on $\pa M$, which can be imposed as a boundary
condition.  One can also  easily show that equations of motion become covariant
ones with respect to Diff$M$ by a simple Weyl rescaling. In fact, this action
is actually Diff$M$-invariant if $\pa M = 0$ because the change is just a total
derivative.

It turns out that the above h-decomposition is not the only one interesting.
There is another important decomposition which we shall call
{\bf g-decomposition}. In the g-decomposition we can in fact define a metric
tensor with respect to SDiff$M$ as we define $g_\mn$ with respect to Diff$M$.
In this case we let $g_\mn = \e^{2\hat\phi}\hat{g}_\mn$ and that
$$\delf g_\mn = \e^{2\hat\phi}\delf \hat{g}_\mn
+ \hat{g}_\mn\delf\e^{2\hat\phi} ,$$
where for some nonvanishing constant $s$
\beqa
\label{exhtg}
\delf\hat{g}_\mn &=& v^\alpha\pa_\alpha\hat{g}_\mn
		+ \hat{g}_{\mu\alpha}\pa_\nu v^\alpha
		+ \hat{g}_{\alpha\nu}\pa_\mu v^\alpha
		- s{\del}_\alpha v^\alpha \hat{g}_\mn , \\
\label{exhtge}
\delf\e^{2\hat\phi} &=& v^\alpha\pa_\alpha\e^{2\hat\phi} +
s\del_\alpha v^\alpha \e^{2\hat\phi}.
\eeqa
Although under Diff$M$ they behave in quite unusual way, but under SDiff$M$
\beqa
\label{esdlie}
\hat{\lie}_v \hat{g}_\mn &:=&\delf\hat{g}_\mn = v^\alpha\pa_\alpha\hat{g}_\mn
		+ \hat{g}_{\mu\alpha}\pa_\nu v^\alpha
		+ \hat{g}_{\alpha\nu}\pa_\mu v^\alpha ,	\\
\label{esdsc}
\hat{\lie}_v\e^{2\hat\phi} &=&\delf\e^{2\hat\phi} =
 v^\alpha\pa_\alpha\e^{2\hat\phi}
\eeqa
so that we can use $\hat{g}_\mn$ to build manifestly SDiff$M$-invariant
actions. Due to the area preserving condition, eq.(\ref{esdlie}) actually
depends on $\hat\phi$. In this case, since a Weyl transformation does
not preserve the structure of eq.(\ref{exhtge}), the action of the form $\CS
(\hat{g})$ is not necessarily Weyl invariant.

Just to make the story complete, we include $g_\mn =
\e^{2\varphi}\tilde{g}_\mn$ such that $\varphi$ transforms like a scalar  and
$\tilde{g}_\mn$ transforms like a tensor under Diff$M$. Then $g_\mn$ and
$\tilde{g}_\mn$ are related by  a conformal diffeomorphism or a Weyl rescaling.
Comparing to the h-decomposition, we can see why fixing $\varphi$ does  not
really fix all the conformal degrees of  freedom. This is because there still
are conformal degrees of freedom in the trace of the metric variation.
We get complete
fixing in the h-decomposition and then the remaining symmetry is SDiff$M$.
Finally, just to summarize, we have the following identity: $g_\mn =
\e^{2\phi}h_\mn = \e^{2\hat\phi}\hat{g}_\mn  = \e^{2\varphi}\tilde{g}_\mn$.

\newsubsection{Symplectic structure of {\rm SDiffM} in 2-d}

Now we can easily check that the volume element
$\hat\mu = e^1\wedge e^2$ is indeed invariant under area preserving
diffeomorphisms in coordinate-independent manner as follows:
\beq
\label{elievol}
\lie_v\hat\mu = \d i_v (e^1\wedge e^2) =
 -*\delta\hat{v} = - \hat\mu\, \delta\hat{v}= 0.
\eeq
This also implies that in 2-d we have a symplectic manifold $(M, \hat\mu)$ and
$v \in \sdiff M$ is nothing but a hamiltonian vector field on $M$.  SDiff$M$ is
a Lie group acting on this symplectic manifold.  Thus locally we have
$$
i_v\hat\mu = -\d H
$$
for some function $H$. $\lie_v H = 0$ implies that SDiff$M$ is a group of
symmetries of this Hamiltonian system. The appearance of such a symplectic
structure is a  unique property in two-dimensions so that there might be an
interesting hamiltonian formalism of 2-d gravity motivated by this.

\newsubsection{Generators}

Note that Diff$S^1$ is related to $\CW_c$ and is not a subgroup of
SDiff$M$. Hence, as far as 2-d gravity is concerned, one can have a hope
that SDiff$M$ may contain information that Diff$S^1$ lacks, but is helpful
to better understand 2-d gravity. Therefore,
explicit forms of generators for sdiff$M$ are much needed to investigate the
representation theory of sdiff$M$, which will reveal many useful properties of
SDiff$M$, as the representations of Diff$S^1$ provide important information to
study conformal field theories, but unfortunately it is a complete mystery.

In general, we are not able to express the generators of the Lie algebra
$\sdiff M$ in terms of a local coordinates explicitly, but the generators
for the abelian subalgebra have local expressions:
\beq
\label{eqsubal}
\left[L_m,\ L_n\right] = 0 \quad{\rm for}\ \
L_n:= x^ny^{-n-1}\pa_x + x^{n-1}y^{-n}\pa_y ,
\eeq
where $(x,y)$ is a set of Riemann normal coordinates. These generators are
well-defined only away from the coordinate origin.

\newsubsection{Variational principle in the h-decomposition}

For a given gravitational action $\CS$ the variation with respect to arbitrary
infinitesimal change of metric is given by
\beq
\label{eqci}
\delta\CS = \int_M \sqrt{|g|}\, T_\mn\delta g^\mn.
\eeq
To derive equations of motion $\delta g^\mn$ is any metric deformation in the
configuration space and $\delta\CS = 0$ is required,  but to derive
stress-energy tensor $\delta g^\mn$ is only along the symmetry directions.
In the latter case, $T_\mn$ can be identified as a stress-energy tensor. Thus
if $\CS$ is diffeomorphism invariant, equations of motion are given by
 $T_\mn = 0$ and $\del^\mu T_\mn = 0$.

But if a theory defined by an action of $\CS(h_\mn,\cdots)$ is not manifestly
invariant under Diff$M$ and the ``metric'' satisfies $h^\mn\delta h_\mn =0$
always under Diff$M$ as in the h-decomposition,  then for equations of motion
we could require
\beq
\label{eqcii}
T^{(h)}_\mn =\half T^{(h)} h_\mn
\eeq
for some function $T^{(h)}$ yet to be determined such that  $\delta\CS =
\half\int_M \sqrt{|h|}\, T^{(h)} h_\mn\delta h^\mn = 0$. Thus it defines a
modified variational principle.  If we regard the LHS of eq.(\ref{eqcii}) as
sort of energy-momentum, it is neither traceless, nor covariantly conserved
unless $T^{(h)}$ is constant such that
\beq
\label{exxii}
\del^{(h)\mu} T^{(h)}_\mn = \half \pa_\nu T^{(h)}.
\eeq
This does not necessarily mean that the theory is ill-defined, but it simply
implies that $h_\mn$ is not a metric. We can always define a new
object\footnote{Such $T_\mn$ also shows up in \cite{jack}.} $T_\mn \equiv
T^{(h)}_\mn -\half T^{(h)} h_\mn$, which is nothing but the stress-energy
tensor with respect to the metric $g_\mn$ and the equation of motion now
reads $T_\mn = 0$.
Thus requiring SDiff$M$-invariance only does not necessarily contradict to the
general covariance.

\newsection{Liouville Gravity}\\[-1cm]

\newsubsection{Generic SDiff$M$ structure}

Liouville gravity is an induced gravity of a two-dimensional system.
For an arbitrary metric before any gauge fixing, the action has a nonlocal form
\beq
\label{eliouv}
\CS_L={c\over 96\pi}\int\!\!\int_M R\Delta^{-1} R
+ {c\over 24\pi}\int_M\Lambda.
\eeq
More precisely,
\beq\label{eliv}
{\CS_{L}={c\over 96\pi}\int_M d^2x_1 d^2x_2
{\sqrt {|g(x_1)|}}{\sqrt {|g(x_2)|}} R(x_2) K(x_1, x_2) R(x_1)
+{c\over 24\pi}\int_M d^2x{\sqrt {|g(x)|}}\Lambda,}
\eeq
where ${\sqrt {|g(x_1)|}}\Delta_1 K(x_1, x_2)=\delta (x_1-x_2)$ and scalar
curvature $R=\Delta\Phi=-g^{\mu\nu}\del_\mu \del_\nu\Phi$ for some $\Phi$
which is not necessarily globally defined.  Note that $K$, or
$\Phi$,  is not uniquely defined but only up to zero modes that are nothing but
harmonic  functions defined on manifold $M$.  Since there is no such a globally
defined nontrivial (\ie not constant) harmonic function on a compact Riemannian
manifold without boundary, one can always express eq.(\ref{eliouv}) locally.
Otherwise, the local action is not uniquely defined. Nevertheless, as we shall
show below, we can interpret this freedom of non-local action as a symmetry of
equations of motion. It is also worth while
to mention that this action is related to
the SL(2,$\IR$)  Chern-Simons theory\cite{verlh} and a complete local form of
this action (up to a surface term)  for the Euclidean signature is given in the
same paper.

If we choose the conformal gauge $g_\mn = \e^{2\phi}\eta_\mn$ for Diff$M$,
$\CS_L$ reduces to the well-known Liouville action. Since $\phi$ transforms
like a scalar, the expression $g_\mn = \e^{2\phi}\eta_\mn$ is covariant
under conformal diffeomorphisms and invariant under isometries.
As a result, the Liouville action is invariant under conformal diffeomorphisms
as well as isometries. Therefore, the conformal
gauge is not a true gauge fixing condition for Diff$M$, but it rather reduces
Diff$M$  to  the space of conformal diffeomorphisms $\CW_c$ and isometries.
Note that the isometry group is part of SDiff$M$. Since the isometry group of
a pseudo-sphere is ISO(1,1) (or ISO(2) for a  sphere) which is isomorphic to
SL(2,$\IR$), this explains the origin of SL(2,$\IR$) symmetry in
\cite{polygrav}.

Normally, we take it for granted that $\phi$ is a scalar under Diff$M$.
However, if $\phi$ does not transform like a scalar, but satisfies
$\lie_v(2\phi) = \del_\mu v^\mu$, we can show that the
expression $g_\mn = \e^{2\phi}\eta_\mn$ is in fact covariant under the general
coordinate transformations, despite $\eta_\mn$ being constant.  Furthermore,
this expression is actually SDiff$M$-invariant because $\lie_v(2\phi) = 0$
for $\del_\mu v^\mu = 0$. Therefore, in this case the conformal
gauge is not a true gauge fixing condition for Diff$M$ and $\eta_\mn$ is not
a constant metric tensor but constant expression for $h_\mn$ in the
h-decomposition.

This also indicates that the quantization of the Liouville gravity
in the conformal gauge should be more involved and perhaps
Batalin-Vilkovisky quantization may be useful due to the secondary gauge
symmetry of $\CW_c$ or $\SDiff M$ that needs to be further fixed\cite{BV}.
(Further work is in progress in this direction.)

In fact we can do more generally. Using any generic decomposition of a metric
$g_\mn = \e^{2\phi}\sigma_\mn$, we can
derive an equivalent Liouville action from eq.(\ref{eliv})
with respect to the background ``metric'' $\sigma_\mn$ as
\beq
\label{esclv}
\CS_L = \CS_{NL}(\sigma) +\CS_{LL}(\sigma, \phi),
\eeq
where
\beqa
\label{esplq1}
\CS_{NL}(\sigma) &=& {c\over 96 \pi} \int_M
\sqrt{|\sigma|}\sqrt{|\sigma|} R(\sigma) K R(\sigma) \\[1mm]
\label{esplq2}
\CS_{LL}(\sigma, \phi) &=& {c\over 24\pi} \int_M d^2x \sqrt{|\sigma|}\left(
\sigma^\mn \pa_\mu\phi \pa_\nu\phi + \phi R(\sigma)
+ \Lambda \e^{2\phi}\right).
\eeqa
If $\Lambda = 0$, $\CS_{LL}(\sigma, h)$ has an additional hidden symmetry that
is not manifest at the action level, but it is a symmetry of the equation of
motion. One can always shift $\phi$ by harmonic functions. This is a reminder
of the fact that the kernel $K$ is only defined up to zero modes that are
harmonic functions. If $\pa M = 0$, then this symmetry corresponds to  shifting
the action merely by a constant. It can be actually shown that $\CS_{LL}$ is a
well-defined integration  over $M$ under the SDiff$M$ gauge symmetry in both
h- and g-decomposition, using equations of motion.

In the h-decomposition, since $h_\mn$ does not really transform like a metric
tensor,  $\CS_{LL}(h, \phi)$ is not manifestly Diff$M$-invariant.   Using
$\del_\mu v^\mu = 0$ and eqs.(\ref{ehtg}-\ref{ehtge}), it can be shown that it
is not SDiff$M$-invariant either in general, unless $v^\alpha\pa_\alpha\phi$ is
harmonic but nonvanishing. At this moment we do not know the effect of
$\del_\mu\del^\mu(v^\alpha\pa_\alpha\phi) = 0$ on SDiff$M$ precisely yet,
so we will just assume that there exist such cases.
(For more discussion, see the last section.) Thus with this constraint
$\CS_{LL}(h, \phi)$ is SDiff$M$-invariant. As a result, $\CS_{NL}$ is also
SDiff$M$-invariant but not Diff$M$-invariant. Note that $\CS_{LL}(h, \phi)$  is
invariant under Weyl rescaling up to equations of motion.

If we want a metric in which the Liouville field transforms like a scalar under
SDiff$M$, then we can take the g-decomposition.
Now the action becomes $\CS_{LL}(\hat{g}, \hat\phi)$ and it is
manifestly SDiff$M$-invariant without imposing any extra condition.
However, in this case we cannot set $\hat{g}_\mn$ to $\eta_\mn$
and there is no conformal symmetry.

\newsubsection{Remarks on $\gamma^\mn\equiv\sqrt{|g|} g^\mn$ case}

In \cite{kmm,jack} it is argued that there exists an action of the
form eq.(\ref{eliouv}) that depends only on the Weyl invariant combination
$\gamma^\mn\equiv\sqrt{|g|} g^\mn$
rather than each separately but differs by local terms.
Here we shall adopt the same idea but we impose the covariant condition
so that the follows are not necessarily the same.
Note that $\gamma^\mn$ is a tensor density so that it is not a metric tensor.
And it does satisfy the criterion for the h-decomposition because
$\gamma^\mn\delf\gamma_\mn = 0$ for any vector field in diff$M$ and
$\delf\sqrt{|g|} = \sqrt{|g|}\del_\mu v^\mu $.
Thus it is a case of the h-decomposition.

In 2-d we have an identity
\beq
\label{eqlv}
\sqrt{|g|} R = R(\gamma) + \Delta^{(\gamma)}\ln\sqrt{|g|},
\eeq
where $R(\gamma)$ is the ``scalar curvature'' computed in terms $\gamma_\mn$
as if $\gamma_\mn$ were a metric tensor and
$\Delta^{(\gamma)} = -\pa_\mu (\gamma^\mn \pa_\nu) = \sqrt{|g|}\Delta$.
This also shows that in general $\sqrt{|g|} R$ is not Weyl-invariant in 2-d
because the second term in the RHS of eq.(\ref{eqlv}) is not Weyl invariant
unless the change $\ln\sqrt{|g|}$ is harmonic.
Locally under coordinate transformations,
$R(\gamma)$ is not invariant but transforms like
\beqa
\label{eqlvi}
\sqrt{|\gamma|}R(\gamma)= R(\gamma) &\to&
J^{-1}\left(R(\gamma) - \Delta^{(\gamma)}\ln J^{-1}\right) ,\\
\sqrt{|\gamma|}\Delta^{(\gamma)}\ln \sqrt{|g|} &\to& J^{-1}
\Delta^{(\gamma)}\ln\left(J^{-1}\sqrt{|g|}\right), \nonumber
\eeqa
where $J$ is the jacobian of the coordinate transformation.
Note that $R(\gamma)$ does not transform like a scalar, unless $J$ is a
harmonic function.

Anyhow, eq.(\ref{esclv}) can be decomposed as
\beqa
\label{eqlvii}
\CS_{L}\!\!\! &=&\!\! \CS_W(\gamma)
+ \CS_{NW}\!\!\left(\gamma, \ln\sqrt{|g|}\right),\\
\CS_W\!\!\! &=&\!\! {c\over 96\pi} \int_M\! d^2x_1 d^2x_2
R(\gamma(x_2)) K(x_1, x_2) R(\gamma(x_2)) ,\\
\CS_{NW}\!\!\left(\gamma, \ln\sqrt{|g|}\right)\!\!\! &=&\!\! {c\over 96\pi}
\int_M\! d^2x \left(\gamma^\mn \pa_\mu\ln\sqrt{|g|} \pa_\nu\ln\sqrt{|g|}
 + 2\ln\sqrt{|g|} R(\gamma) + 4\Lambda \sqrt{|g|}\right)\! ,\quad\
\eeqa
where $\CS_W$ is Weyl invariant, but the integrand is not a scalar density.

Nevertheless, we can check how $\CS_W$ behaves under SDiff$M$. In fact, we can
take a short cut, using the  generic property of h-decomposed metric instead of
directly computing the Lie derivative of $\CS_W$. One can easily show that
$\CS_L$ and $\CS_{NW}$ are  SDiff$M$-invariant if
$v^\alpha\pa_\alpha\ln\sqrt{|g|}$ is
harmonic so that $\CS_W$ has to be SDiff$M$-invariant. If $\Lambda = 0$,
$\CS_{NW}$ is also Weyl invariant up to equations of motion.

The rationale behind this approach is that the combined symmetry of area
preserving diffeomorphisms and Weyl rescaling provides the same amount of gauge
degrees of freedom as diffeomorphism invariance. This is because we can trade
off $\CW_c\in\Diff M$ with Weyl rescalings. To realize this idea one need to
show that any change of coordinate basis can be achieved by combined coordinate
transformation by SDiff$M$ and Weyl rescaling so that the integration of
$\CS_W$ can be consistently defined over $M$.  If we impose the covariant
condition $\del_\mu v^\mu = 0$, this can be accomplished with the help of
equations of motion.

\newsection{Geometric Liouville Gravity}\\[-10mm]

\newsubsection{Lagrangian and Duality}

For argument's sake we shall start from the dual form of Liouville  gravity
action, then later decompose into the geometric Liouville action written in
terms of SDiff$M$ variables in the g-decomposition to recover the results in
\cite{glg}. To obtain the result in the h-decomposition we simply redefine the
variables. However,  equations of motion will have a different form due to the
different variational principle.

For this purpose, we enlarge the configuration space to the
space of all metrics for arbitrary genus surfaces and take Diff$M$ as gauge
symmetry. Following \cite{glg}, we relate the scalar field $\Phi$ such that
$\Delta\Phi=R$ to the affine spin connection $\omega$ by
\beq
\label{edual}
\omega= -*\d\Phi \quad {\rm or}\quad
\omega_\mu = \epsilon_\mn\pa^\nu\Phi .
\eeq
In this sense, $\Phi$ and $\omega$ are dual to each other.

Now $\delta\omega = 0$ is satisfied so that the curvature two-form can be
written as
\beq\label{ecur}
\hat{R} = (\d + \delta)\omega =\triangle *\Phi,
\eeq
where $\triangle$ is the Laplace-Beltrami operator.
Then the first term in eq.(\ref{eliouv}) can be rewritten as
\beq
\label{elcl}
\int_M \triangle^{-1}\hat{R}\wedge * \hat R
=\int_M (\d+\delta)^{-1}\omega\wedge *(\d+\delta)\omega.
\eeq

Using the Hodge dual property of scalar product on compact oriented manifold,
now the theory of action eq.(\ref{eliouv}) is locally equivalent to that of
an action which can be written as a local form
\beq\label{egliouv}
\CS_A={c\over 96\pi}\int_M\omega\wedge *\omega + {c\over 24\pi}\int_M *\Lambda
= {c\over 96\pi}
\int_Md^2x {\bf e}\left(\omega_\mu\omega^\mu+4\Lambda\right)
\eeq
with gauge fixing conditions
\beqa\label{eapgaugea}
\delta\omega\!&\!=\!&\! -\del_\mu\omega^\mu=0 ,\\
\label{eapgaugeb}
n^\mu w_\mu &\!=\!& 0 \quad {\rm on}\ \pa M\ {\rm for}\ n^\mu\perp\pa M,
\eeqa
where ${\bf e} = \sqrt{|g|}$ is the zweibein volume element.
This gauge fixing condition, which is motivated by the area preserving
condition of $\sdiff M$, fixes the local Lorentz symmetry in the frame space,
that is, SO(1,1) (or SO(2) depending on the signature) in this case.
Due to this gauge fixing,
the ambiguity related to the zero modes of $\Delta$ in the original
Liouville action $\CS_L$ is no longer present because $\CS_A$ is now invariant
under $\omega \to \omega + \lambda_H$, where one-form $\lambda_H$ corresponds
to the zero mode shift of $\Phi$. Thus, $\CS_A$ is a uniquely defined
local action corresponding to the nonlocal action $\CS_L$ and the zero mode
ambiguity is now identified as a symmetry.

Notice that this
gauge fixing condition is invariant under Diff$M$ and can be imposed
globally on $M$, although $\omega$ itself is not globally defined on $M$.
Thus the metric is still not constrained.
Metric gauge fixing should be imposed independently. In the frame space,
although $\del^\mu\omega_\mu$ is invariant under only global SO(1,1) (or
SO(2)),  but in fact $\del^\mu\omega_\mu = 0$ is preserved under global
GL(2,${\IR}$).  Under local SO(1,1), the gauge fixing condition is preserved
only up to a harmonic function. Note that local Lorentz transformations and
Diff$M$ do not commute. In general, $\CS_A$ in eq.(\ref{egliouv}) is more
general than the original Liouville gravity because the equivalence to the
Liouville gravity action is true only if $\delta\omega = 0$.
In some sense this is due to the
nonlocality of the original action. Later we shall abandon this gauge fixing
condition and investigate $\CS_A$ itself.

For the time being, zweibeins $e_\mu^a$ will be considered to be the
only fundamental variables, rather than treating $(e,\omega)$ independently as
is often done in a gauge theory formulation of gravity.  Thus the affine
spin connection  is always computed in terms of zweibeins, so we do not need
to impose the torsion-free condition as a constraint. In this sense, the
action takes a form analogous to gauge theory action $\int F^2$.

Under Diff$M$ the Lagrangian in eq.(\ref{egliouv}) changes according to the
Lie derivative as
\beq
\label{eviii}
\lie_v (\omega\wedge *\omega)
=\d\left(i_v\omega\wedge *\omega -\omega\wedge i_v *\omega \right).
\eeq
Although it is a total derivative, if $\pa M\neq 0$, this does not necessarily
vanish when integrated over $M$. But it vanishes if $v^\alpha \in \sdiff M$.

To compare to the Liouville case, we rewrite in terms of $\Phi$ in
eq.(\ref{edual}),  then the action eq.(\ref{egliouv}) reads
\beq
\label{eix}
\CS_A = {c\over 96\pi}\int_M d^2x \sqrt{|g|}
\left[g^\mn\pa_\mu\Phi\pa_\nu\Phi +
\lambda\left( R-\Delta\Phi\right) +4\Lambda \right] ,
\eeq
where $\lambda$ is the Lagrange multiplier for the constraint $R = \Delta\Phi$
and $\Phi$ is defined only up to a harmonic function. Of course, this action
can also be derived directly from eq.(\ref{eliouv}) but then the zero mode
ambiguity may not be clearly resolved.

Variation with respect to $\Phi$ leads to
$$
\lambda = 2 \Phi
$$
so that we can rewrite the action for this value as
\beq
\label{egleq}
\CS_A= {c\over 48\pi}\int_M d^2x \sqrt{|g|}
\left( -\half g^\mn\pa_\mu\Phi\pa_\nu\Phi +\Phi R + 2\Lambda \right),
\eeq
which leads to equivalent equations of motion. Here, the Liouville field
behaves like a real scalar field except that the kinetic energy has an opposite
sign compared to eq.(\ref{eix}).

Note that there is no exponential potential term compared to the usual
Liouville action so that $\Phi$ is not the usual Liouville field. To identify
the usual Liouville field the metric needs to be decomposed. To reproduce the
result in \cite{glg} we take the g-decomposition, which produces manifestly
SDiff$M$ covariant objects. Then $\CS_A$ decomposes as
\beqa
\label{eqglac1}
\CS_A &=& \CS_G(\hat{g}) + \CS_{AL}(\hat{g}, \hat\phi),  \\[1mm]
\label{eqglac2}
\CS_G(\hat{g}) &=& {c\over 48\pi}\int_M d^2x \sqrt{|\hat{g}|}
\left( -\half \hat{g}^\mn\pa_\mu\Phi(\hat{g})\pa_\nu\Phi(\hat{g}) +
\Phi(\hat{g}) R(\hat{g}) \right), \\[1mm]
\label{eqglac3}
\CS_{AL}(\hat{g}, \hat\phi) &=& {c\over 24\pi}\int_M d^2x \sqrt{|\hat{g}|}
\left(  \hat{g}^\mn\pa_\mu\hat{\phi}\pa_\nu\hat{\phi} +
\hat{\phi} R(\hat{g}) +\Lambda\e^{2\hat\phi}\right).
\eeqa
$\CS_G(\hat{g})$ is equivalent to the geometric (Liouville)  action in
\cite{glg}, whilst $\CS_{AL}(\hat{g},\hat\phi)$ is the usual Liouville action
in this metric decomposition.  Note that $\CS_G$ is manifestly
SDiff$M$-invariant, but not Diff$M$-invariant because $\hat\phi$ is absent. As
a result, $\CS_{AL}$ is not Diff$M$-invariant but SDiff$M$-invariant. In the
h-decomposition, $\CS_G(h)$ is Weyl invariant under any  Weyl transformation of
the metric $g_\mn$ because $h_\mn$ does not change but only $\phi$
changes. But in the g-decomposition $\CS_G(\hat{g})$ is not
Weyl invariant in the usual sense. However, there is an analogous symmetry
for $\CS_G(\hat{g})$ under $\Phi(\hat{g}) \to \Phi(\hat{g}) + \delta\Phi$,
which leaves $\CS_G(\hat{g})$ invariant up to equations of motion. This is
due to the zero mode ambiguity of the original action.

One can easily observe that in fact $\CS_G$ contains all the classical
information about $\CS_A$ for $\Lambda = 0$. Since the tree level cosmological
constant is not really important  toward quantum theory in this context, it is
good enough to use $\CS_G$ as a whole classical action.  $\CS_{AL}$ simply
generalizes to include $\Lambda\neq 0$. This is  why $\CS_G$ is often just
ignored by imposing a constraint on $\hat\Phi$ and $\CS_{AL}$ is used to
describe the Liouville theory in the third kind of decomposition $g_\mn =
\e^{2\varphi}\tilde{g}_\mn$, in which $\tilde{g}_\mn$  transforms like a metric
tensor.  Here, we now realize that this is just one special case and we can fix
Diff$M$ in many different ways. We can select either $\CS_G$ or  $\CS_{AL}$ to
describe the classical theory in any metric decomposition.

In the g-decomposition, $\CS_G$ is equivalent to the geometric action
constructed in \cite{glg}. If we represent the action in terms of $\omega$,
then the action is an analog of the vorticity hamiltonian in fluid dynamics,
where the coupling constant $c$ takes the role of the density and
$\omega^\mu\pa_\mu \in \sdiff M$.

In the g-decomposition, the
stress-energy tensor works the same way as in Diff$M$ case.
But, to derive a conserved quantity in the h-decomposition, now we should use
the modified variational principle with respect to Diff$M$.
Then from eq.(\ref{eqglac2}) we obtain
\beq
\label{exxi}
T^{(h)}_\mn =\pa_\mu\Phi(h)\pa_\nu\Phi(h)
-\half h_\mn  h^{\alpha\beta}\pa_\alpha\Phi(h)\pa_\beta\Phi(h)
+\half h_\mn\Delta^{(h)}\Phi(h) =\half T^{(h)} h_\mn,
\eeq
where $h^\mn T^{(h)}_\mn =T^{(h)}$ determines
$T^{(h)}=\Delta^{(h)}\Phi(h)= R(h)$.
Thus $T^{(h)}_\mn$ is neither traceless nor covariantly conserved.
As we alluded before, we can always define
$T^{{\rm phys}}_\mn \equiv T^{(h)}_\mn - \half R(h)h_\mn = 0$,
which implies an equivalent result in the Diff$M$ case by a simple
Weyl transformation and vice versa.
Therefore, classically it does not contradict to the general covariance.

On the other hand, we can also regard $T$ in the above as a gauge parameter
such that $T= R(h)$ is not an identity but an equation of motion.
Fixing $T$ corresponds to fixing the gauge degrees of freedom of
Diff$M$. For example, $T=0=R(h)$ leads to the SDiff$M$ invariance of
$\CS_{AL}(h, \phi)$.

In the g-decomposition, $T$ must vanish, leaving $R(\hat{g}) = 0$. Then $R(g) =
-2\Lambda$. Since $\hat{g}_\mn$ behaves like a metric tensor under SDiff$M$, we
can imitate the Diff$M$ case to solve this equation.  Nonetheless, it still
leaves two components of $g_\mn$ undetermined, namely, $\hat\phi$ and one of
$\hat{g}_\mn$. In some cases, $\hat\phi$ and the remaining component of
$\hat{g}_\mn$ may not be separable in $g_\mn$ in practice. One good example is
$\hat{g}_\mn = \e^{2\hat\rho}\eta_\mn$ gauge fixing. This does not necessarily
indicate quantization in the g-decomposition will be the same as the Diff$M$
case because the transformation law for $\hat{g}_\mn$ is not really independent
from $\hat\phi$ due to the area preserving condition.

\newsubsection{More about $\delta\omega = 0$}

Since this gauge fixing of local Lorentz invariance is the key to relate
$\CS_A$ to $\CS_L$, it deserves some more attention. We already pointed out
that this gauge fixing is necessary due to the nonlocality of $\CS_L$.

First, we may attempt to find if there is any equivalent gauge fixing condition
acting directly on zweibein themselves. Using
$$
[\del_\alpha, \del_\nu]\  e_\mu^a = -e_\lambda^a R^\lambda_{\ \mu\alpha\nu},
$$
we obtain
\beqa
\label{eqi}
\del_\mu\omega^{\mu a}_{\ \ b} &=& -E^\nu_b\del^\mu\del_\mu e_\nu^a \nonu
&=& -E^\nu_b\del_\nu\del^\mu e_\mu^a -\half\delta^a_{\ b}R
-E^\nu_b\del^\mu \left(\del_\mu e_\nu^a -\del_\nu e_\mu^a \right).
\eeqa
Since $a\neq b$, $R$-term vanishes. Now one may be tempted to conclude that
$\del^\mu e_\mu^a = 0$ and $\del_\mu e_\nu^a - \del_\nu e_\mu^a = 0$ seem to be
sufficient conditions to satisfy $\del_\mu\omega^\mu = 0$, but this is more
than what we need for gauge fixing. In fact the above are nothing but  $\delta
e^a = 0 = \d e^a$ so that $e^a$ becomes a harmonic one-form. Using the torsion
constraint, we obtain $\omega^a_{\ b}\wedge e^b = 0$. In this case one can
easily see that $\d e^a=0 $ implies $\omega_\mu = 0$ and that
$g^\mn\omega_\mu\omega_\nu = 0$,  \ie\ the action itself vanishes. Thus it does
not seem to be possible to  obtain any simple condition directly on zweibeins.

There is another way to check this gauge fixing condition. A zweibein basis of
a frame changes under Diff$M$, so we need to understand the global property of
such a gauge fixing more carefully. Since $\delta\omega=0$, locally we can
obtain $\omega = \delta W$ for some two-form
$W = \half W_\mn \d x^\mu\d x^\nu$. Let $W_\mn  = \epsilon_\mn\Phi$, then
\beq
\label{eqii}
\omega = \delta\hat\Phi ,
\eeq
where $\hat\Phi \equiv *\Phi$.
Comparing to the curvature two-form $\hat R = \d\omega$, we obtain
\beq
\label{eqiii}
\hat R = \d\delta\hat\Phi .
\eeq
Globally $\omega$ is determined up to co-closed one-form $\lambda$ such
that $\delta\lambda = 0$, hence $\omega = \delta\hat\Phi + \lambda$.
Eq.(\ref{eqiii}) in turn implies $\lambda$ is also a closed one-form.
so that $\lambda$ is in fact a harmonic one-form. Therefore, $\delta\omega = 0$
gauge is equivalent to choosing $\lambda$ to be a harmonic one-form.

Now let us check if $\delta\omega = 0$ uniquely fixes all the gauge degrees of
freedom or there are any secondary gauge degrees of freedom which leave
$\lambda$ invariant. From eq.(\ref{eqiii}) we can always add an exact one-form
$\d F$, which makes the complete decomposition of $\omega = \delta\hat\Phi
+\lambda_H + \d F$, where $\lambda_H$ denotes a harmonic one-form. The gauge
fixing condition implies $\Delta F = 0$ so that $F$ must be a harmonic function
and $dF$ is a harmonic one-form. Thus in general $\delta\omega = 0$ does not
fix $\lambda$ uniquely. In other words, the gauge fixing condition is preserved
by the change of  Laplacian of a harmonic function\footnote{Such a situation
happens in QED too.  The Lorentz gauge  condition is invariant if the gauge
parameter is a harmonic function.}.  In the case of $g_\mn =
\e^{2\rho}\eta_\mn$, one can easily show that this actually corresponds to
local SO(1,1) (or SO(2)) symmetry.

On a compact Riemannian manifold without boundary $\delta\omega = 0$
does fix the gauge completely (up to a discrete set of harmonic one-forms
depending on the topology) because the above decomposition is unique
according to the Hodge decomposition theorem\cite{cnh}, and that a harmonic
function is necessarily a constant so that $\d F = 0$. Thus $\omega =
\delta\hat\Phi + \lambda_H$ is the unique decomposition on a compact manifold
without boundary.  The symmetry corresponding to $F$ is no longer local, but
global SO(1,1) (or SO(2)) symmetry.

\newsubsection{$(e,\omega)$ independently}

One can also think about
\beq\label{eqxi}
\CS_A(e,\omega)=c\int_M\omega\wedge *\omega + c\Lambda\int_M e^1\wedge e^2
\eeq
as a defining action for a gravitational theory. Compared to the first-order
formalism, there is no explicit derivative terms in this action because
$\omega$ is no longer directly related to zweibeins.  Thus in particular there
is no explicit kinetic energy term. We only have ``mass'' terms, if we regard
$(e,\omega)$ as gauge fields of ISO(1,1) (or SL(2,$\IR$)). In particular,
classical equations of motion are just $\omega = 0$ and $\Lambda = 0$
which can be regarded as unbroken phase of a gravitational theory {\it \`a la}
Witten\cite{wit21d} because the classical action vanishes for these values and
there are no dynamical degrees of freedom.
This indicates that perhaps $\CS_A(e,\omega)$ may define an integrable theory,
if not topological. The integrability of this theory may not be surprising at
the end because after all it is related to the Liouville theory which is
integrable.

Here, in principle $\omega^\mu$ can be any vector on $M$. The reason why
SDiff$M$ becomes relevant in \cite{glg} is because $\delta \omega = 0$ is
imposed as a constraint so that $\omega^\mu\in \sdiff M$ as a vector field on
$M$. Then $\CS_A$ becomes a hamiltonian which describes how $M$ deforms keeping
the area of $M$ fixed. Once $\delta\omega = 0$ is imposed, this is dual to
the usual Liouville action as we constructed and there is an
additional ``Weyl'' symmetry: $\omega \to \omega + *\d\rho$ for some function
$\rho$. But this constraint is not essential in general for $\CS_A$. However,
to define a reasonable gravitational theory on
a Riemannian manifold $M$, there is one constraint we must impose.
This is the torsion-free constraint $\d\e + \omega\e = 0$ and it also
identifies $\omega$ as the affine spin connection.

Thus imposing the following constraint, we recover the dynamics:
\beq
\label{eqxii}
\CD_\mu e_\nu^a := \pa_\mu e_\nu^a -\Gamma^\lambda_{\mn} e_\lambda^a
+\omega_{\mu\ b}^a e^b_\nu = 0,
\eeq
where $\CD_\mu$ is the covariant derivative acting both on $M$ and
on the frame space. This constraint recovers the previous relation between
spin connection and zweibeins, eq.(\ref{espc}).

If we introduce only the torsion-free constraint into the action, we obtain
\beq
\CS_A(e,\omega)=c\int_M\left(\omega\wedge *\omega
+\lambda (\d\e +\omega\wedge\e)	\right)
+ c\Lambda\int_M e^1\wedge e^2.
\eeq
Variation with respect to $\omega$ derives $\lambda \e = -*\omega$ and the rest
of equations of motion are simply
$$\omega=0,\ \ \d\e=0,\ \ ,\Lambda=0.$$
Compared to the Liouville case in which $R=-2\Lambda$, we obtain a different
result, unless $\Lambda = 0$. This is why we need $\delta\omega = 0$ for
$\CS_A$ to be related to the Liouville gravity.

\newsection{Conclusion and Discussions}

We have provided a general framework to construct SDiff$M$-invariant
gravitational theories in two-dimensions, which are not necessarily manifestly
Diff$M$-invariant. From Diff$M$'s point of view, fundamental field variables
are no longer globally defined, which is not unusual in gauge theories.  Two
different ways of defining such field variables are introduced: h- and
g-decomposition. In the h-decomposition, it is necessary to impose equations of
motion to define a consistent integration on $M$, whilst the integration in the
g-decomposition is well-defined without imposing any conditions as in Diff$M$
cases if an integrand is a scalar density with respect to SDiff$M$.

So far as Liouville gravity is concerned, we have shown that there is SDiff$M$
invariant subsystem which contains sufficient information about the original
system. In this sense it does not violate the general covariance  at the level
of classical equations of motion. Now, one may ask if there are  any merits to
use SDiff$M$ invariant system rather than Diff$M$ invariant  system, since they
come out to be equivalent classically. Nevertheless, we expect a real
difference may show up in quantum theories, particularly in which any dilaton
degrees of freedom are completely frozen. To describe a physical system we are
required to fix all the gauge degrees of freedom so that in principle we can
allow a physical gravitational system in which  all conformal degrees of
freedom in Diff$M$ are spontaneously broken as well as Weyl symmetry. And the
``physical'' dilaton may incorporate not only Goldstone modes of Weyl symmetry
but also those of conformal symmetry in Diff$M$. In other words, we can define
a (massive) dilaton as a (pseudo-)Goldstone  boson of Weyl$\otimes$Diff$M$ to
SDiff$M$ symmetry breaking. To describe quantum physics of such a dilaton, the
formalism we described in this paper should be useful. So we expect that the
key to resolve the mystery of the massive dilaton may reside in this framework.

Also SDiff$M$ invariance provides a framework to describe intrinsically
a theory defined on a manifold with boundary without introducing a boundary
term. This inevitably addresses an issue of the energy-momentum conservation
at the boundary, but as we pointed out the physical energy-momentum can always
be defined to be conserved.

Many questions remain to be answered. For example,
in the g-decomposition SDiff$M$ invariance is manifest by construction, whilst
in the h-decomposition we need an extra constraint to show SDiff$M$ invariance
in the Liouville case. It is not clear how this extra constraint
$\del^\mu\del_\mu v^\alpha\pa_\alpha\phi = 0$ restricts SDiff$M$. In the
simpler case of isometry group, this condition requires $R(h)\propto e^{2\phi}$
so that we can anticipate that it may actually restrict the form of $h_\mn$.
In other words, the Liouville action is not a good candidate to be SDiff$M$
invariant in the h-decomposition. It would be interesting to know if there is a
modified action that is invariant under SDiff$M$ without any further
constraint.

It is also necessary to know how to quantize such a SDiff$M$ invariant system
consistently. Our hope is that there may be a generation of dilaton potential
in this approach because there is no symmetry which prohibits this from
happening.  From the conventional point of view,  we can speculate that the
trace of the graviton may be  absorbed into the dilaton to  provide dilaton
mass. We hope further investigation in this direction reveals more physical
roles of the area preserving diffeomorphism in gravity in general.



\begin{flushleft}
{\bf Acknowledgements}:\\[3mm]
\end{flushleft}
We thank P. Nelson for communications, R. Jackiw and B. Zwiebach
for reading the manuscript.

{\renewcommand{\Large}{\large}

}

\end{document}